# Toward a Robust and Generalizable Metamaterial Foundation Model

*Namjung Kim[*], Dongseok Lee, Jongbin Yu, Sung Woong Cho, Dosung Lee, Yesol Park, and Youngjoon Hong[*]*
[*] Corresponding authors

Jongbin Yu, Dosung Lee, Yesol Park, Prof. N. Kim

Department of Mechanical Engineering, Gachon University, Seongnam 13120, Republic of Korea

Dongseok Lee, Sung Woong Cho

Department of Mathematical Sciences, Korea Advanced Institute of Science and Technology, Daejeon, 34141, Republic of Korea

Prof. Y. Hong

Department of Mathematical Sciences, Seoul National University, Seoul, 08826, Republic of Korea

Interdisciplinary Program in Artificial Intelligence, Seoul National University, Seoul, 08826, Republic of Korea

Keywords: Foundation model, In-context learning, Metamaterials, Inverse design, Bayesian transformer

## Abstract

Advances in material functionalities drive innovations across various fields, where metamaterials—defined by structure rather than composition—are leading the way. Despite the rise of artificial intelligence (AI)-driven design strategies, their impact is limited by task-specific retraining, poor out-of-distribution (OOD) generalization, and the need for separate models for forward and inverse design. To address these limitations, we introduce the Metamaterial Foundation Model (MetaFO), a Bayesian transformer-based foundation model inspired by large language models. MetaFO learns the underlying mechanics of metamaterials, enabling probabilistic, zero-shot predictions across diverse, unseen combinations of material properties and structural responses. It also excels in nonlinear inverse design, even under OOD conditions. By treating metamaterials as an operator that maps material properties to structural responses, MetaFO uncovers intricate structure-property relationships and significantly expands the design space. This scalable and generalizable framework marks a paradigm shift in AI-driven metamaterial discovery, paving the way for next-generation innovations.

# Introduction

Recent breakthroughs in additive manufacturing (AM) have accelerated the design and realization of metamaterials, which has helped overcome the limitations of traditional fabrication. High-precision AM technologies now enable the construction of complex, multiscale architectures[1–4], a development that is driving innovations in diverse domains such as robotics[5,6], medicine[7], energy[8], pharmaceutics[9], and aerospace engineering[10,11]. These architected materials have exhibited unprecedented optical[12,13], mechanical[14,15], thermal[16–18], electromagnetic, and acoustic[19,20] behaviors and are driving a paradigm shift in material science. Although conventional approaches, such as topology optimization, heuristic algorithms, and bio-inspired strategies, have facilitated significant advancements in metamaterials, their scalability and adaptability remain limited, particularly as functional requirements become more intricate[21]. Thus, the complexity of modern metamaterials necessitates more sophisticated design methodologies beyond conventional approaches. Addressing these challenges demands a shift toward artificial intelligence (AI)-driven frameworks.

AI-driven design methodologies are redefining the landscape of metamaterial design, offering novel means to efficiently navigate vast, high-dimensional design spaces[22]. These techniques establish direct correlations between microstructural topology and effective material properties across multiple scales, which significantly accelerates both forward and inverse design with remarkable precision. AI has already demonstrated success in optimizing multi-material[23], multi-scale[24], and multi-functional metamaterials[25]. However, existing AI-driven methods still face fundamental barriers: (1) extensive retraining is necessary when applied to different metamaterial families or property objectives, (2) predictive accuracy is limited for unseen configurations or operating ranges, (3) vulnerability to noise in training data, and (4) reliance on separate models for forward and inverse design tasks. To overcome these limitations, a transformative approach is required that enhances generalization, strengthens out-of-distribution (OOD) predictions, reduces sensitivity to noisy data, and unifies the metamaterial design process through recent AI advancements.

To this end, we introduce the Metamaterial Foundation Operator (MetaFO)—a foundation model that extends the transformative paradigm of large language models (LLMs) into the domain of mechanical metamaterials. Foundation models have demonstrated remarkable success across a wide range of disciplines, including natural language processing, drug discovery[26], pathology[27], and molecular modeling[28], by leveraging general-purpose architectures trained on vast datasets. Among these, LLMs stand out as the most prominent success story, largely due to their scalability, cross-domain adaptability, and exceptional generalization capabilities. Inspired by the architectural and training strategies of LLMs, recent works have begun to explore neural operator-based approaches for learning physical laws and solving complex PDE-governed systems[29]. However, the application of such foundation models to data-driven design and prediction tasks in metamaterials remains largely underexplored. MetaFO addresses this gap by pioneering a physics-aware, few-shot generalizable operator framework for forward prediction, inverse design, and uncertainty quantification in metamaterials, which lays the groundwork for a new class of mechanics-driven foundation models inspired by neural operator architectures for metamaterial design[30–32].

The key idea of MetaFO is that all different metamaterial systems, despite their geometrical diversity,

share fundamental underlying principles of mechanics. Mechanics, the study of forces and their effects on matter, provides a unifying framework for understanding and engineering material behavior across multiple scales. Its core principles, including stress, deformation, and energy transfer, govern a wide range of physical systems, including the mechanical behavior of metamaterials. By learning a unified mapping between material properties and structural responses for each metamaterial, MetaFO captures these commonalities as a neural operator. This approach eliminates the need for expensive retraining processes and enables the model to extrapolate across metamaterials with different graph geometries. By leveraging this shared operator framework, MetaFO drastically reduces computational overhead and data requirements, allowing a single neural network to perform forward and inverse design tasks across multiple metamaterial systems. Inspired by recent advances in in-context learning (ICL)[33–35], MetaFO achieves zero-shot inference[36,37] by dynamically adapting to new tasks through user prompts without requiring task-specific retraining. This capability enables MetaFO to predict metamaterial responses for previously unseen configurations and base materials with exceptional accuracy, seamlessly integrating interpolation, extrapolation, inpainting, uncertainty reduction, and inverse design as downstream tasks. By leveraging fine-tuning, MetaFO refines its predictions across diverse scenarios, ensuring robust performance even in highly nonlinear and data-scarce regimes.

To the best of our knowledge, MetaFO represents the first application of this paradigm in material science, establishing a new standard for AI-driven metamaterial design. To enhance predictive robustness, MetaFO integrates a Bayesian transformer architecture with prior-data fitted networks (PDFNs)[38] and leverages prior knowledge from extensive metamaterial databases to capture intricate structure-property relationships. By defining prior distributions over unseen material configurations, MetaFO achieves robust, high-fidelity predictions even in highly nonlinear regimes. This breakthrough enables MetaFO not only to predict novel metamaterial properties but also to reveal previously uncharted correlations between base materials and their responses, laying the groundwork for the next generation of intelligent metamaterial design.

## Results

**Metamaterial Foundation Operator (MetaFO)**

Building on prior work that explored ICL for physical phenomena[39] and AI-driven metamaterial design methodologies[40,41], we introduce MetaFO, a transformative approach that expands the boundaries of metamaterial learning and inverse design. By representing unit cells as operators, it learns a generalizable mapping from material properties to structural responses (Fig. 1a). This self-supervised approach, trained on a diverse dataset, allows it to handle complexities such as missing or noisy data without task-specific retraining, addressing four key challenges: forward prediction, extrapolation, noise reduction, and inverse design.

Unlike conventional AI-based design strategies, which are typically trained on specific datasets with specific materials or unit cell configurations, MetaFO is designed to operate across multiple datasets, comprising various materials as well as unit cell configurations. This enables MetaFO to learn universal mechanical principles, resulting in broad generalization across materials and configurations of metamaterials. Instead of merely predicting outputs for individual test samples, MetaFO constructs a foundational latent space that captures the fundamental relationships between material properties and unit cell responses. This enables zero-shot and few-shot learning for novel metamaterial configurations, significantly enhancing efficiency in real-world design applications.

Before deployment, MetaFO undergoes a single pretraining phase on thousands of synthetically generated datasets, allowing it to learn the core physics underlying metamaterial behavior across both elastic and plastic regimes. Once trained, MetaFO serves as a general-purpose foundation model for metamaterial behavior prediction and design, capable of addressing four critical challenges in the field—including forward prediction, performance extrapolation, uncertainty quantification, and inverse design, as outlined in Fig. 1—without requiring additional retraining. This scalability and adaptability mark a significant leap toward truly data-efficient, physics-aware AI for metamaterial discovery.

The training dataset is constructed from 10 basis unit cells and their linear combinations, used to construct a comprehensive database of mechanical metamaterials. Each unit cell is represented as a graph, defined by nodes and connecting edges, enabling a compact and flexible encoding of topology. This graph-based representation offers a balance of computational efficiency and design versatility, facilitating the generation of diverse metamaterial architectures. The selected 10 unit cells (Fig. 1b and Fig. S1) serve as structural primitives, from which a total of $2^{10} - 1 = 1,023$ unique configurations are generated through all possible non-empty combinations. By further incorporating 10 distinctive base materials, the dataset expands to 10,230 unique metamaterial structures. The corresponding mechanical responses of each unit cell, as well as base materials (Fig. S2), are computed using an in-house workflow integrated with the commercial finite element software ABAQUS. Additional methodological details can be found in the Methods section and Supporting Information.

MetaFO's transformer-based architecture is trained to directly map material properties to the compressive responses of each unit cell (Fig. 1c). This one-time pretraining step equips MetaFO with a generalized learning framework that eliminates the need for case-specific model retraining and ensures adaptability across diverse metamaterial configurations. Once trained, MetaFO can be directly applied to four real-world scenarios: 1) Task 1: predicting the compressive behavior of unseen metamaterials; 2) Task 2: predicting the compressive behavior of metamaterials with missing data, both inside and outside the design range; 3) Task 3: handling noisy data; and 4) Task 4: performing inverse design to generate metamaterials with a target property. All tasks are summarized in the illustrative schematics shown in Fig. 1c, which outline the corresponding inputs and outputs for each stage.

**Task 1: Generalization to unseen material–property combinations**

The growing use of mechanical metamaterials across various fields calls for a deeper understanding of their behavior under both elastic and plastic deformation regimes in full three-dimensional structures. However, existing predictive models are often limited to two-dimensional, pixelated structures or purely elastic responses, making them less applicable to real-world systems. Advancing metamaterial design requires high-fidelity models that can accurately capture the behavior of three-dimensional architectures in both linear and nonlinear regimes.

In this context, we demonstrate the ability of MetaFO to generalize to unseen material–property combinations. Fig. 2a illustrates the MetaFO prediction pipeline. Given input pairs consisting of material properties ($\varepsilon_{m_i}, \sigma_{m_i}$) and corresponding unit cell responses ($\varepsilon_{u_j}, \sigma_{u_j}^{m_i}$), MetaFO predicts the unit cell response for either an unseen configuration ($\sigma_{u_{unseen}}^{m_i}$) or an unseen material ($\sigma_{u_j}^{m_{unseen}}$). Here, $\varepsilon_{m_i}$ and $\sigma_{m_i}$ denote the strain and stress of material $m_i$, where $i$=1, … ,10 indicates different material types. Similarly, $\varepsilon_{u_j}$ and $\sigma_{u_j}^{m_i}$ represent the strain and stress of unit cell $u_j$, composed of material $m_i$, where $j$=1, …, 10 indicates different unit cell configurations. Strain is defined as the ratio of deformed to original length, and stress refers to the applied force per unit area. Fig. 2b presents MetaFO's performance across a representative set of metamaterials with varying geometrical complexities, capturing the stress–strain relationships of various combinations of basis unit cells. The results show that MetaFO achieves exceptional predictive accuracy, consistently capturing both linear and nonlinear mechanical responses across diverse unit cell configurations. Additional cases are provided in the Supporting Information. Notably, MetaFO accurately captures abrupt transitions between linear and nonlinear regimes. In doing so, it addresses a long-standing challenge in metamaterial modeling.

Fig. 2c presents the zero-shot prediction accuracy of MetaFO in comparison with several well-established machine learning (ML) models. As shown, MetaFO outperforms the other models by a factor of 10 to 100, demonstrating its superior prediction capability. Even in a zero-shot setting, where no prior training examples of the target structure are provided, MetaFO maintains high predictive accuracy, a fact that highlights its ability to infer underlying physical principles directly from context. Beyond standard interpolation, MetaFO exhibits remarkable generalization to previously unseen unit cell structures, as quantified in Fig. 2d. MetaFO's predictions accurately capture key nonlinear mechanical responses—including the abrupt stress drop after yielding, the plateau region arising from strut bending, and the densification behavior at high strains—even for previously unseen unit cell configurations. Fig. 2e illustrates the variation in prediction accuracy as the number of combined unit cells increases from one to four and includes unseen configurations. The prediction accuracy might reduce when the complexity of unit cell combinations increases. However, remarkably, all predictions remain within a 2.5% error range—even for unseen unit cells—which highlights the strong generalizability of MetaFO. The elevated response at 0.1 strain is attributed to initial instability during the early stage of loading. Similar to LLMs, the inclusion of additional prompt inputs enhances predictive accuracy. This trend is validated in Fig. 2f, where performance improvement nearly doubles as the number of prompts increases from one to three. However, beyond three prompt sets, the marginal gains diminish. This behavior underscores MetaFO's superiority and adaptability over conventional ML and DL models.

**Task 2: Robustness to missing data in the base material property function**

Data obtained from experiments often suffers from data loss due to low-precision measurement systems and experimental uncertainties. This issue is particularly pronounced near the yield point or in extrapolated regions, where mechanical behavior becomes increasingly nonlinear. Ensuring robust handling of missing data is therefore critical to enhancing a model's predictive reliability. MetaFO directly addresses this challenge by learning to infer missing material properties across two key domains: interpolation (within-range predictions) and extrapolation (beyond observed data).

Fig. 3a provides an overview of MetaFO's missing data prediction framework. Given input pairs consisting of material properties $(\varepsilon_{m_i}, \sigma_{m_i})$ and corresponding unit cell responses $(\varepsilon_{u_j}, \sigma_{u_j}^{m_i})$ with data loss, MetaFO predicts the unit cell response $(\sigma_{u_j}^{m_i})$ even in the presence of missing data—either in the middle of the input vector (interpolation) or at its end (extrapolation). The regions with missing data are highlighted in green and orange, corresponding to interpolation and extrapolation tasks, respectively. Fig. 3b shows the quantification of its interpolation accuracy in the strain range of 0.1 to 0.3, where material property data is deliberately withheld to evaluate its predictive performance. This strain range is particularly critical as it often encompasses the yield point, where uncertainty peaks due to the abrupt transition in mechanical behavior. Even in these absent regions, MetaFO reconstructs full mechanical response curves with high precision, maintaining consistency across diverse metamaterial configurations. To further assess robustness, we examine four representative unit cells, each tested across four distinct base materials. Despite significant variations in stress–strain nonlinearity, MetaFO consistently produces accurate reconstructions, demonstrating its strong generalization across mechanical systems.

The extrapolation of data presents a fundamentally greater challenge than interpolation due to its unbounded nature. Fig. 3d evaluates MetaFO's extrapolation performance, where missing data occurs between strain values of 0.3 and 0.5. Despite the inherent difficulty of predicting beyond the training range, MetaFO successfully captures the compressive behavior of all tested unit cells and materials, even under strong material and geometrical nonlinearities. Notably, its prediction accuracy remains remarkably high, reinforcing its ability to operate reliably in previously unseen design spaces. Fig. 3c quantitatively evaluates MetaFO's predictive performance across different levels of structural complexity and strain. As the number of unit cell combinations increases, the prediction error systematically decreases. This trend arises because more complex combinations are associated with significantly larger training datasets. For instance, the number of samples for single-unit combinations is $\binom{10}{1} = 10$. Meanwhile, for four-unit combinations, it is $\binom{10}{4} = 210$, yielding a dataset that is 21 times larger. The richer data availability enables MetaFO to learn more robust representations, thereby enhancing prediction accuracy under increased structural complexity and strain. Remarkably, even in the most challenging cases, interpolation and extrapolation errors remain below 2%. This result underscores MetaFO's exceptional reliability in missing data scenarios. We also highlight that the predictive accuracies in both interpolation and extrapolation scenarios surpass those of previous ML methods, as shown in Fig. 2c. These results establish MetaFO as a highly robust and generalizable predictive framework, capable of accurately reconstructing mechanical responses despite missing data. By maintaining high accuracy under nonlinear conditions, MetaFO sets a new benchmark for data-driven metamaterial design, ensuring greater reliability and adaptability across diverse applications.

**Task 3: Overcoming the limit of prediction uncertainty from noised data**

The accuracy of predictive models is fundamentally constrained by the precision of input data; this constraint makes robustness to measurement noise a critical requirement for practical applications. In compression experiments, noise arises due to instrumentation limitations and experimental uncertainties, particularly when the deformation involves bending or local crack propagation in materials, which is very common in mechanical metamaterials. This phenomenon can amplify measurement deviations. In addition, since high-precision measurement systems are costly, experimental outputs are often inherently limited by available resources, necessitating a model that can reliably infer accurate predictions even under noisy conditions.

In this task, we evaluate the resilience of MetaFO to data uncertainty, demonstrating its ability to effectively suppress noise while preserving high-fidelity predictions. The data uncertainties considered in this study are depicted in the database and prompt, as shown in Figs. 4a (i) and (ii), respectively. A Gaussian noise-induced measurement error is introduced into the material properties ($\sigma_{m_i}$) as well as unit cell responses ($\sigma_{u_j}^{m_i}$). Conventional models would amplify this uncertainty and thereby propagate noise through the unit cell's response. In contrast, MetaFO accurately reconstructs compressive behavior and yields a near-exact match between predicted and actual curves under 5% Gaussian noise—an ability further validated across four distinct unit cell configurations, as shown in Fig. 4b. The figure demonstrates the noise reduction capability of MetaFO at various noise levels.

To systematically evaluate noise robustness, Fig. 4c presents MetaFO's predictive accuracy as the noise level in the prompt data increases from 3% to 10%, across varying numbers of prompt pairs. Here, each prompt pair consists of a material property and its corresponding unit cell response. The number of such pairs directly influences the model's performance. Remarkably, even when noise levels reach 10% and the number of prompt pairs increases from 3 to 5, the resulting prediction error remains below 4%—lower than the injected noise itself. As the number of prompt pairs increases, the variance associated with unit cell geometries decreases, which indicates improved prediction confidence. This level of robustness notably exceeds that of conventional ML models, which tend to degrade more rapidly under similar noise conditions. Importantly, this robustness is observed not only for unseen materials but also for previously untested unit cell geometries, highlighting MetaFO's strong generalization across diverse metamaterial systems under uncertainty.

Fig. 4d investigates the impact of the proportion of noisy data in the database on prediction accuracy. As the ratio of noisy data increases, prediction error rises across all noise levels. However, the degradation remains marginal up to 30% contamination, highlighting the robustness of MetaFO. This suggests that MetaFO can maintain reliable performance even when up to 30% of the database is corrupted. Fig. 4e further examines the influence of noisy and clean database volumes on prediction accuracy. Surprisingly, incorporating 20% and 40% noisy data into the original dataset results in approximately 10% and 20% improvements in prediction accuracy, respectively. This suggests that augmenting the dataset—even with noisy samples—can enhance the performance of MetaFO, potentially by promoting generalization. In practical scenarios, such resilience is particularly advantageous in distributed data collection environments where measurement precision varies. By effectively mitigating the adverse effects of experimental noise, MetaFO sets a new benchmark for AI-driven metamaterial design that ensures reliable and precise predictions even under uncertain or imperfect data. These findings underscore MetaFO's potential to accelerate innovation in metamaterial design by reducing reliance on high-precision instrumentation and expanding the feasibility of real-world experimental workflows.

**Task 4: Unified framework for forward and inverse design of mechanical metamaterials**

A long-standing challenge in the field of mechanical metamaterials is the disjoint treatment of the forward and inverse design processes. Traditional approaches often require separate models or computational pipelines to predict a unit cell's behavior from a given structure (forward prediction) and to infer optimal unit cells for target properties (inverse design). This separation introduces inefficiencies, redundant training, and increased computational costs, particularly when scaling to large design spaces. To address this limitation, we propose a unified framework that seamlessly integrates forward and inverse tasks within a single operator-based model architecture, MetaFO.

Fig. 5a illustrates the inverse design framework of MetaFO. The input consists of the vectorized coordinates of $n$ nodes, which serve as the design space for unit cell generation. Upon receiving user-specified target pairs—each comprising desired material properties and their corresponding unit cell responses—MetaFO generates an $n \times n$ matrix $A$, which encodes the predicted connectivity information among the nodes. As the entries in matrix $A$ are continuous values indicating the likelihood of connections, a thresholding technique is applied to convert them into a binary adjacency matrix, where edge presence is denoted by 1 and absence by 0. Fig. 5b presents representative inverse design examples with varying unit cell complexities and numbers of prompts. In each case, the top two plots show the input material properties and the corresponding target responses. The number of such property-response pairs reflects the number of user-defined prompts for each task. The bottom two plots display the predicted and ground-truth connectivity graphs, demonstrating that MetaFO consistently recovers accurate node connectivity across all test cases.

This inverse design task holds critical importance in the development of mechanical metamaterials, as it directly enables the discovery of microstructures tailored to exhibit user-defined properties. Unlike forward prediction, which merely evaluates the behavior of known structures, inverse design empowers engineers and researchers to actively navigate vast and complex design spaces to achieve functional goals. This capability is especially vital in applications requiring on-demand customization, such as biomedical implants, aerospace components, and soft robotics, where performance specifications must be met under strict constraints. By efficiently generating candidate structures that satisfy target responses, inverse design accelerates the innovation cycle, reduces reliance on trial-and-error prototyping, and unlocks new possibilities in the rational design of next-generation metamaterials.

## Discussions

MetaFO represents a significant advancement in the development of general-purpose AI frameworks for mechanical metamaterials. By incorporating physics-aware priors and leveraging a transformer-based architecture, MetaFO transcends the limitations of traditional task-specific models and sets a new standard for generalizability, robustness, and adaptability in the design and prediction of complex material systems.

A key innovation of MetaFO lies in its operator-based formulation, which enables seamless learning across diverse unit cell geometries and material compositions. This formulation acknowledges the shared physical laws governing metamaterial responses, allowing the model to encode generalized mappings from input material properties to output structural behavior. As demonstrated across four representative tasks—forward prediction, missing data prediction, noise resilience, and inverse design—MetaFO consistently delivers accurate results, even under challenging conditions such as unseen geometries, extrapolated inputs, and measurement uncertainties. Notably, it achieves high fidelity in both elastic and plastic regimes, which overcomes a long-standing barrier in modeling mechanical metamaterials.

In contrast to conventional deep learning strategies that require extensive retraining and manual feature engineering, MetaFO's one-time pretraining on synthetically generated datasets enables it to operate in a zero- or few-shot setting. This significantly reduces the cost and time associated with simulation-driven workflows and paves the way to interactive, prompt-based design—an emerging paradigm inspired by the success of LLMs. The capacity of MetaFO to generalize across 10,000+ diverse metamaterial configurations without retraining underscores its potential as a foundation model for mechanics, in the same way that LLMs have transformed language and vision tasks.

The unified architecture for forward and inverse tasks further consolidates MetaFO's role as a next-generation design engine. Rather than relying on separate pipelines, MetaFO employs a single neural operator framework capable of both evaluating existing structures and generating new ones to meet user-defined mechanical targets. This design unification not only reduces computational redundancy but also enables efficient, targeted exploration of high-dimensional design spaces. The model's ability to reconstruct microstructural connectivity directly from desired responses, even under sparse prompt conditions, marks a major advancement in inverse design capabilities.

Beyond its technical performance, MetaFO's resilience to noise and missing data makes it particularly suitable for real-world deployment, where data quality is often compromised. The model's robustness, even under 10% Gaussian noise and 30% corrupted training data, significantly lowers the barrier for experimental integration. This property is especially valuable in decentralized, resource-constrained settings—such as in situ testing, wearable sensors, or distributed manufacturing environments—where high-precision instrumentation may be unavailable.

## Methods

### Data preparation

To construct the basis unit cells, a $5 \times 5 \times 5$ grid of control nodes is defined within a dimensionless design space. Edges connecting these nodes form the geometry of each unit cell, which is represented as a graph. This graph-based representation balances geometric expressiveness and computational efficiency and thereby enables the generation of a broad spectrum of metamaterial designs—from simple cubic forms to complex, interconnected architectures. A set of 10 graph-based unit cells with cubic symmetry was selected as the basis set (see Fig. S1 in the Supporting Information). By taking the linear combinations of these 10 basis unit cells, a total of 1,023 unique geometries were generated. Each geometry was paired with 10 different material property sets, yielding a dataset of 10,230 microstructures in total.

To evaluate their mechanical performance, the compressive behavior of each unit cell was simulated using the finite element method (FEM). Numerical simulations were conducted in ABAQUS, a commercial software developed by Dassault Systèmes Simulia Corp., with an in-house Python script for automation. Quasi-static conditions were imposed to minimize inertial effects, and rigid plates were positioned at the top and bottom of each structure to replicate experimental compression tests. The bottom plate remained fixed, while the top plate applied a prescribed strain rate. To eliminate undesired boundary artifacts, periodic boundary conditions were applied to the remaining faces of the unit cell. From each simulation, 21 equally spaced data points were extracted from the resulting stress and strain vectors to represent the mechanical response. The detailed simulation settings are described in the Supporting Information.

**Operator learning framework**

The metamaterial response function (strain–stress curve, SS-curve) is governed by both the material property (material SS-curve) and the unit cell geometry. When the geometry of the unit cell $u$ is fixed and only the material property changes, the response of the unit cell becomes solely dependent on the material property. In this paper, we aim to approximate the response function of a unit cell for a new material property by using prior data from material–response pairs with the shared geometry $u$. To this end, we develop a learning-based operator model, MetaFO, which directly maps material property functions to their corresponding unit cell response functions. Given a spatially varying material property $f_m(\mathbf{x})$ over a domain $\Omega$, MetaFO predicts the corresponding response function $f_u^m(\mathbf{x})$ without solving governing equations. MetaFO learns the operator $\mathcal{F}$ directly from data, mapping function pairs $(f_m, f_u^m)$ without additional training for each new input. This operator captures spatial dependencies and geometric features, allowing MetaFO to generalize to previously unseen material properties or unseen prompts (i.e., unit cell geometry). Moreover, MetaFO robustly interpolates and extrapolates response functions across diverse material configurations.

**Schematic flow of MetaFO**

To approximate the mapping between material property functions and their corresponding unit cell response functions, we require a computational framework capable of directly handling function-to-function relationships. Since these functions are defined over continuous domains and contain infinitely many values, directly representing the function pairs $(f_m, f_u^m)$ is computationally intractable. A practical solution is to discretize the domain into a finite set of grid points and represent each function by its values at these points, which forms a vector approximation. As the number of grid points increases, this vector approximation more accurately captures the underlying characteristics of the function. In this work, we leverage this discretization strategy to effectively approximate the relationship between material properties and their corresponding responses for fixed unit cell geometries. Instead of mapping directly between continuous functions, we employ a neural network, MetaFO, which learns the mapping between discretized function vectors and thus effectively models the underlying operator, as shown in Fig. 6.

Given a fixed unit cell geometry $u$, MetaFO predicts the metamaterial response functions $f_u^{m_{k+1}}$ from the corresponding material property functions $f_{m_{k+1}}$ of an unseen material $m_{k+1}$. We assume access to $k$ known material–response pairs, each comprising $T$ strain observations. These pairs, denoted as $\left\{\left(f_{m_i}, f_u^{m_i}\right)\right\}_{i=1}^{k}$,

are discretized as
$$\{\varepsilon_{m_i}, \sigma_{m_i}, \varepsilon_u, \sigma_u^{m_i}\}_{i=1}^k \in \mathbb{R}^{T \times 4},$$
where $(\varepsilon_{m_i}, \varepsilon_u)$ and $(\sigma_{m_i}, \sigma_u^{m_i})$ represent the strain and stress values of the material $m_i$ and the metamaterial with fixed geometry $u$, respectively. For inference, MetaFO receives strain–stress data from a new material, along with query strain values for the unit cell. This input is denoted as $\{\varepsilon_{m_{k+1}}, \sigma_{m_{k+1}}, \varepsilon_u\} \in \mathbb{R}^{T \times 3}$, from which the model predicts the corresponding metamaterial response $\sigma_u^{m_{k+1}}$.

The MetaFO algorithm proceeds through the following steps. First, two independent multilayer perceptron (MLP) encoders are employed to process the prompt. A domain encoder, $\phi_{\text{domain}}$, maps each three-dimensional domain token $\mathbf{x}_{\text{domain}} = \{\varepsilon_{m_i}, \sigma_{m_i}, \varepsilon_u\}_{i=1}^k \in \mathbb{R}^{T \times 3}$ to a $d_{\text{inp}}$-dimensional embedding $\mathbf{z}_{\text{domain}}$. Simultaneously, a solution encoder, $\phi_{\text{sol}}$, maps each scalar solution token $\mathbf{x}_{\text{sol}} = \{\sigma_u^{m_i}\}_{i=1}^k \in \mathbb{R}^{T \times 1}$ to $\mathbf{z}_{\text{sol}}$ in the same $d_{\text{inp}}$-dimensional space. The embeddings $\mathbf{z}_{\text{domain}}$ and $\mathbf{z}_{\text{sol}}$ are concatenated to form a unified representation
$$\mathbf{z} = \mathbf{z}_{\text{domain}} \oplus \mathbf{z}_{\text{sol}} \in \mathbb{R}^{2d_{\text{inp}}}$$
for each token. These $kT$ encoded tokens are assembled into a tensor
$$\mathbf{E}_{\text{known}} \in \mathbb{R}^{k \times T \times 2d_{\text{inp}}}$$
by flattening over both the token and function dimensions. This tensor is then processed by a transformer encoder with multi-head self-attention, producing the contextualized prompt representation $\mathbf{H}_{\text{enc}} \in \mathbb{R}^{(kT) \times 2d_{\text{inp}}}$. Next, MetaFO encodes the input of $n$ unknown queries, given by
$$X_{\text{domain}}^{\text{unknown}} = \{\varepsilon_{m_{k+i}}, \sigma_{m_{k+i}}, \varepsilon_u\}_{i=1}^n \in \mathbb{R}^{T \times 3}$$
through the domain encoder to obtain $\mathbf{Z}_{\text{domain}}^{\text{unknown}} \in \mathbb{R}^{n \times T \times d_{\text{inp}}}$, where $n$ is the number of target material queries for which the model predicts the metamaterial responses $\{\sigma_u^{m_{k+1}}\}_{i=1}^n$. To incorporate global context, MetaFO computes a context vector $\mathbf{c}_{\text{global}} \in \mathbb{R}^{T \times d_{\text{inp}}}$ by averaging the solution embeddings of the known responses. This global vector is then broadcast and concatenated with the unknown domain tokens to form a consistent dimensional input.

To enhance the representation, the architecture incorporates a context-conditioned noise injection mechanism. Specifically, a learnable noise bank $\mathbf{E}_{\text{noise}} \in \mathbb{R}^{K \times d_{\text{inp}}}$ and a noise query network $\psi$ are introduced. The network $\psi$ takes as input a broadcast version of the global context vector $\mathbf{c}_{\text{global}}$ and produces dynamic weights
$$\mathbf{w} = \psi\left(\text{Expand}(\mathbf{c}_{\text{global}})\right) \in \mathbb{R}^{(nT) \times K}.$$
These weights are used to generate dynamic noise embeddings via weighted summation over the noise bank, $\mathbf{n}_{\text{dynamic}} = \mathbf{w} \cdot \mathbf{E}_{\text{noise}} \in \mathbb{R}^{(nT) \times d_{\text{inp}}}$. The final decoder input $\mathbf{E}_{\text{dec}} \in \mathbb{R}^{(nT) \times 2d_{\text{inp}}}$ is obtained by concatenating the unknown domain embeddings with the corresponding dynamic noise embeddings. This sequence is processed by a transformer decoder, which uses the encoder output $\mathbf{H}_{\text{dec}}$ as cross-attention memory. The decoder produces a contextualized representation $\mathbf{H}_{\text{dec}} \in \mathbb{R}^{(nT) \times 2d_{\text{inp}}}$. Finally, a multilayer perceptron maps $\mathbf{H}_{\text{dec}}$ to the target output dimension, yielding predicted values $\mathbf{Y} \in \mathbb{R}^{(nT)}$, which are reshaped to $\mathbb{R}^{n \times T}$, to represent the final metamaterial response functions. The complete algorithm is summarized in Algorithm 1.

**Theoretical basis of MetaFO**

In this work, we also address the setting in which the dataset includes observational noise. Under such conditions, the discrepancy between the predicted and true response functions for a given prompt can be more appropriately quantified using the Kullback–Leibler (KL) divergence rather than a deterministic mean squared error (MSE) loss. Assuming the observational noise is Gaussian and independent across function values $f_{u_j}^{m_i}(\varepsilon_{u_j,t})$, minimizing the MSE is mathematically equivalent to minimizing the KL divergence. This perspective allows us to interpret MetaFO as a posterior predictive approximator within a probabilistic inference framework.

*Prompt-based inference*

During inference, MetaFO follows a prompt-based retrieval and prediction strategy. Given a query material function, the model refers to the prompt set and uses learned attention mechanisms to identify relevant patterns and generate the corresponding response function. Attention weights are computed based on the similarity between the query representation and the prompt embeddings, enabling MetaFO to generalize to unseen material configurations through interpolation and extrapolation beyond the support of the training distribution. The operator is parameterized by trainable neural network weights $\theta$, which include all components of the encoder, decoder, and attention layers. The prior knowledge encoded through transformer-based pretraining allows MetaFO to rapidly integrate newly observed material functions and produce robust, zero-shot predictions without requiring any additional fine-tuning.

*Posterior inference over function operators*

In the MetaFO setting, each observation is a pair of functions $\left(f_{m_i}(\varepsilon_{m_i,t}), f_u^{m_i}(\varepsilon_{u,t})\right)$, where $f_{m_i}$ represents the material function and $f_u^{m_i}$ the corresponding response function, representing material properties and corresponding system responses, respectively. The space $Q$ then generalizes to $\mathcal{O}$, the space of all admissible operators mapping between function spaces. MetaFO approximates these operators by learning from large collections of such function pairs, leveraging its graph-based and attention-driven neural architecture to encode prior structure and adaptively update through data-driven posterior learning.

The learning objective is to identify an operator $o \in \mathcal{O}$ that maps $f_{m_i}$ to $f_u^{m_i}$, drawn from a prior $\Pi(o)$ defined over the space of operators $\mathcal{O}$. Given a dataset of function pairs $\left\{\left(f_{m_i}, f_u^{m_i}\right)\right\}_{i=1}^{k}$, the likelihood of observing these mappings under the operator $o$ is expressed as

$$L\left(f_u^{m_i} \mid f_{m_i}; o\right)$$

The likelihood function $L$ quantifies how well the operator $o$ explains the observed functional relationship between the material function and the response function. In other words, $L$ is the operator-level likelihood function that measures the consistency of the mapping from $f_{m_i}$ to $f_u^{m_i}$ under the operator $o$. The posterior distribution over operators is therefore given by

$$\Pi\left(o \mid \{(f_{m_i}, f_u^{m_i})\}_{i=1}^k\right) = \frac{\Pi(o) \prod_{i=1}^k L(f_u^{m_i} \mid f_{m_i}; o)}{\int_{\mathcal{O}} \Pi(o') \prod_{i=1}^k L(f_u^{m_i} \mid f_{m_i}; o') do'}$$

However, the infinite-dimensional nature of $\mathcal{O}$ renders direct integration intractable. MetaFO approximates this posterior by drawing samples $\{o_j\}_{j=1}^N$ from the prior $\Pi(o)$ or from an appropriate proposal distribution and employing Monte Carlo estimation

$$\int_{\mathcal{O}} \Pi(o') \prod_{i=1}^k L(f_u^{m_i} \mid f_{m_i}; o') do' \approx \frac{1}{N} \sum_{j=1}^N \prod_{i=1}^k L(f_u^{m_i} \mid f_{m_i}; o_j)$$

*Transformer-based posterior predictive approximation*

In principle, Bayesian operator learning aims to infer the full posterior distribution over operators. However, owing to the intractability of directly sampling from infinite-dimensional posterior spaces, MetaFO approximates the posterior predictive operator using a transformer-based neural network trained to minimize the discrepancy between predicted and observed response functions. This training process can be viewed as performing approximate maximum a posteriori (MAP) estimation, where the model parameters converge toward the mode of the posterior distribution. Furthermore, the attention mechanism and learned function priors within MetaFO serve as implicit approximations of posterior predictive inference, enabling robust generalization to unseen material–property combinations.

Given a prompt $\mathcal{P} = \{(f_{m_i}, f_u^{m_i})\}_{i=1}^k$, which encodes known material–response mappings, MetaFO uses a transformer-based neural network with parameters $\theta$ to approximate the posterior predictive operator. Specifically, the model learns a conditional predictive distribution $p_\theta(f_u^{m_i} \mid f_{m_i}, \mathcal{P})$, which serves as an implicit approximation to the true Bayesian posterior predictive distribution $\Pi(f_u^{m_i} \mid f_{m_i})$. For any query material function $f_{m_*}$, the transformer-based predictive distribution aims to remain close to the ideal Bayesian posterior predictive. This objective is formalized by minimizing the expected KL divergence between the learned predictive distribution and the true posterior predictive distribution across all possible material functions:

$$\mathbb{E}_{f_m}\left[KL\left(p_\theta(\cdot \mid f_m, \mathcal{P}) \| \Pi(\cdot \mid f_m)\right)\right] < \epsilon,$$

for an arbitrarily small $\epsilon > 0$. In this formulation, the predictive distribution $p_\theta$ reflects the output of the trained transformer model, while the attention mechanism and learned function priors within MetaFO serve as a structural approximation to posterior predictive inference. In practice, this corresponds to learning an operator that closely approximates the mode of the posterior distribution (MAP estimation), with the transformer's architecture enabling efficient generalization and context-aware predictions for new material configurations.

## Data availability

All data used and generated in this study are available in the article and Supplementary Information. Additional data related to this study are available from the corresponding author upon request.

## Code availability

The software needed to reproduce the results are available from the corresponding author upon request.

## Acknowledgments

The work of Y.H. was supported by the Basic Science Research Program through the National Research Foundation of Korea (NRF) funded by the Ministry of Education (NRF-2021R1A2C1093579), and by the Korean government (MSIT) (RS-2023-00219980). The work of N.K. was supported by the National Research Foundation of Korea (NRF) grant funded by the Korean government (MSIT) (No. 2022R1C1C1009387).

## Author contributions

Namjung Kim and Youngjoon Hong designed the research; Dongseok Lee, Jongbin Yu, Sung Woong Cho, and Dosung Lee performed the research; Namjung Kim, Dongseok Lee, Jongbin Yu, Sung Woong Cho, Dosung Lee, Yesol Park, and Youngjoon Hong analyzed the data; and Namjung Kim and Youngjoon Hong wrote the paper.

## Competing interests

The authors declare no competing interests.

# Figures

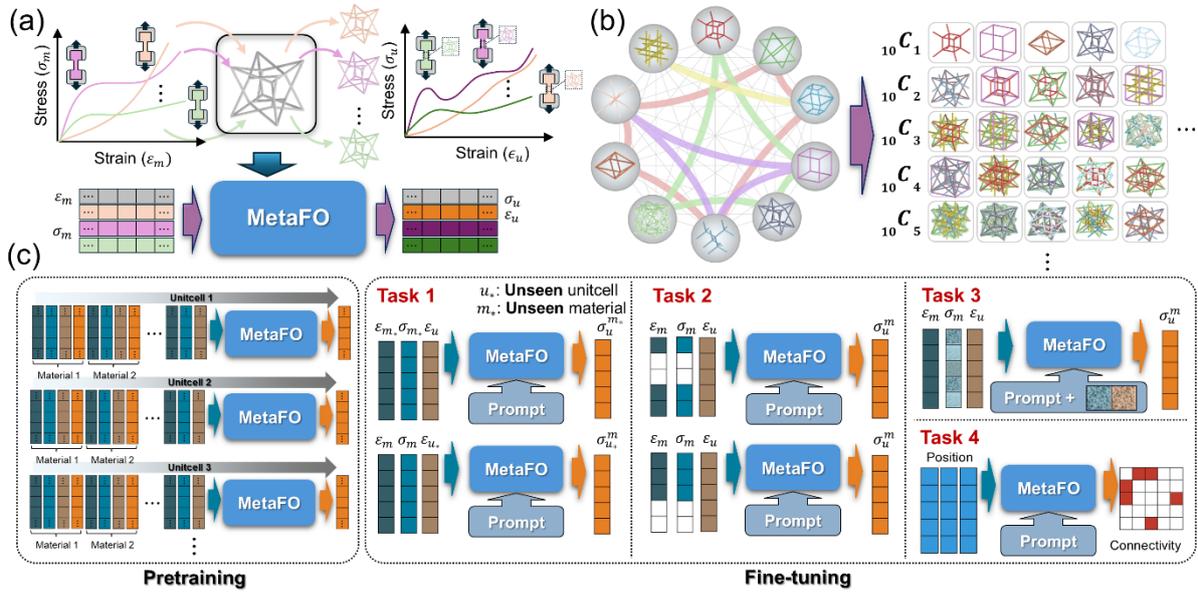

**Figure 1. MetaFO: a generalizable operator-based framework for metamaterials.**

(a) MetaFO models each unit cell as an operator that maps material properties to structural responses.

(b) Ten basis unit cells are used to generate 1,023 topologies, combinatorially expanded with 10 base materials.

(c) MetaFO's transformer-based architecture is trained once and applied to four tasks: generalization, missing data, noise robustness, and inverse design.

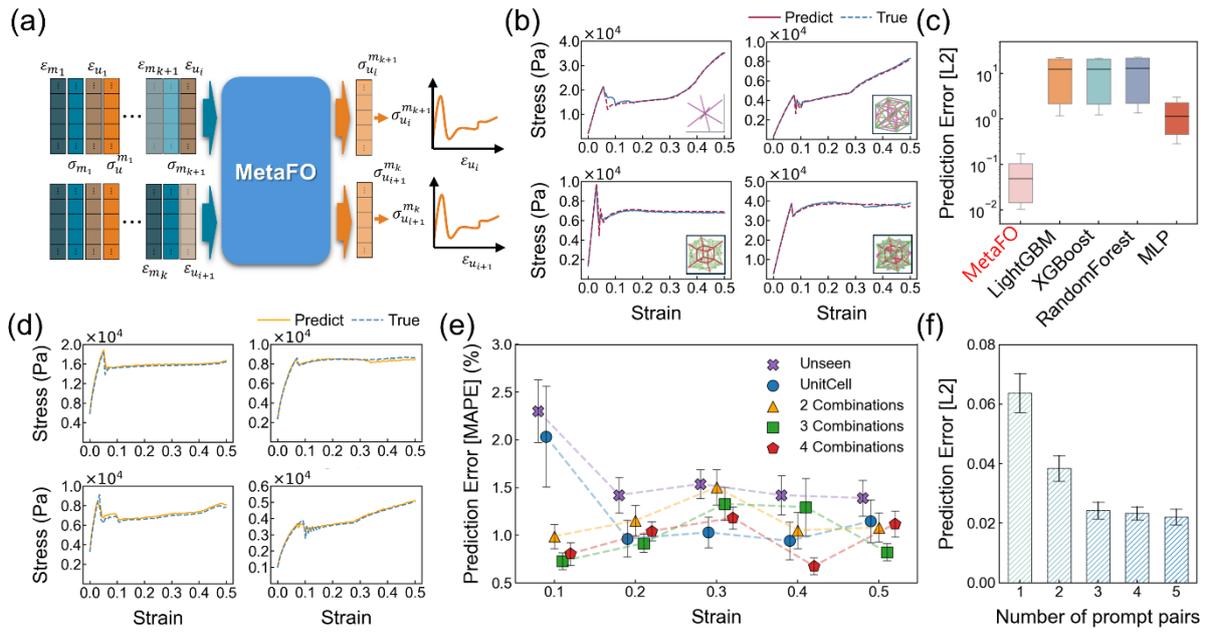

**Figure 2. Generalization performance of MetaFO to unseen material properties and unit cell configurations.**

(a) MetaFO prediction pipeline extrapolates stress–strain responses from known material–structure pairs to unseen materials or unit cells.

(b) Predicted responses across various geometries show high agreement with ground truth, capturing both linear and nonlinear behaviors.

(c) Comparison of zero-shot prediction errors across machine learning models shows that MetaFO outperforms baselines by 10–100×.

(d) MetaFO captures key nonlinear mechanics, including yielding, plateauing, and densification, even for unseen structures.

(e) Prediction error remains below 3% as unit cell combinations increase in complexity.

(f) Prompt-based accuracy improves with more contextual inputs, although diminishing returns are observed beyond three prompts.

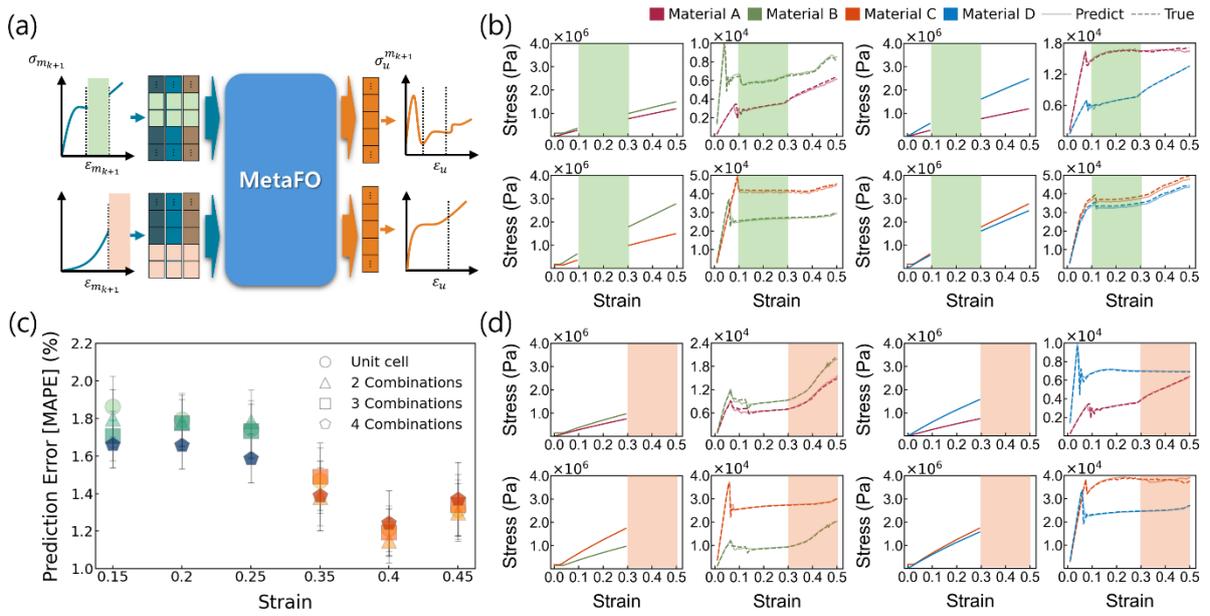

**Figure 3. Missing data prediction performance of MetaFO.**

(a) Schematic of MetaFO's framework for predicting mechanical responses under missing data conditions. Green and orange regions indicate interpolation and extrapolation zones, respectively.

(b) Interpolation accuracy in the 0.1–0.3 strain range captures transitions around yield points, despite withheld material data.

(c) Quantified prediction errors across increasing structural complexity; all cases maintain <4% error for both interpolation and extrapolation.

(d) Extrapolation performance beyond the training range (0.3–0.5 strain) shows accurate predictions across nonlinear behaviors.

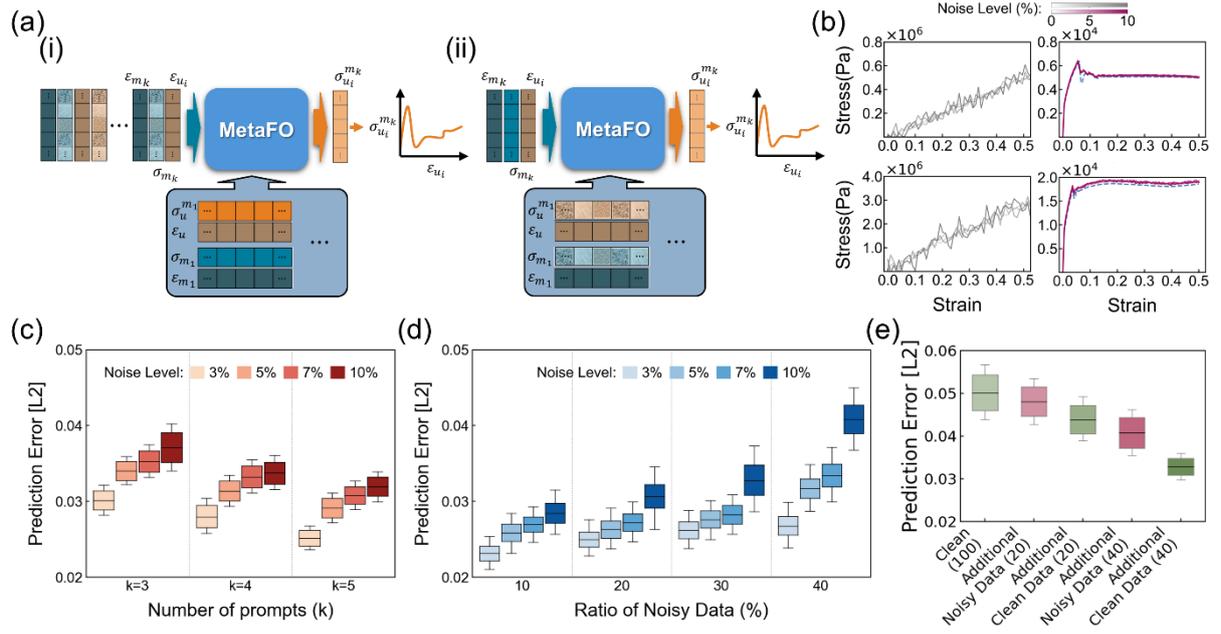

**Figure 4. Robustness of MetaFO to data uncertainty and measurement noise.**

(a) Noise injection settings: (i) within the database and (ii) within prompt inputs, affecting both material and structural responses.

(b) Predicted compressive behavior under 5% Gaussian noise shows strong agreement with ground truth across four unit cell types.

(c) Prediction accuracy under increasing prompt noise (3–10%) and prompt counts (3–5); MetaFO maintains low error despite noise escalation.

(d) Paradoxical performance gain occurs when augmenting the training set with 20–40% noisy data, demonstrating resilience to low-quality datasets.

(e) Adding 20-40% noisy data to the original dataset improves accuracy by up to 20%, indicating that even noisy samples can enhance MetaFO's predictions.

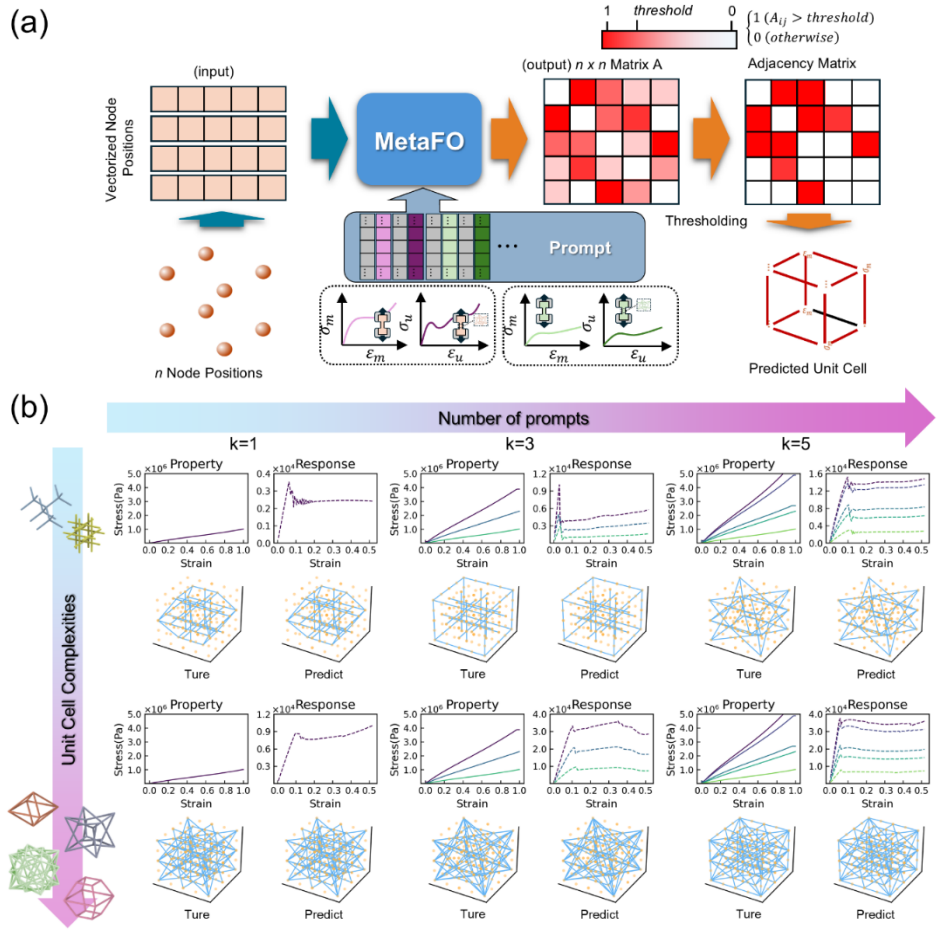

**Figure 5. Inverse design of metamaterial unit cells using MetaFO.**

(a) Inverse design pipeline: given target material–response pairs, MetaFO predicts an n×n connectivity matrix defining unit cell topology. Thresholding converts predicted probabilities to binary adjacency graphs.

(b) Representative design results across varying complexities. Top: input properties and target responses. Bottom: predicted and ground-truth graphs show accurate recovery of node connectivity.

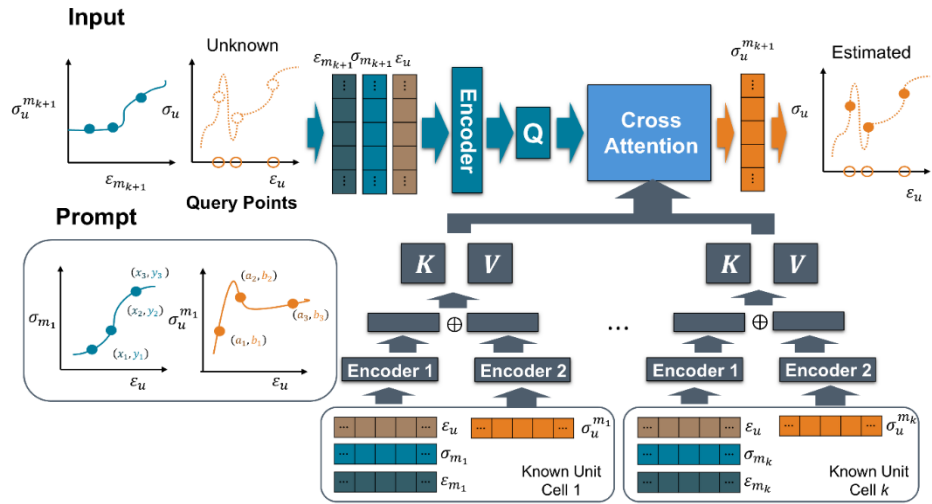

**Figure 6. Operator learning via discretized function mapping in MetaFO.**

MetaFO approximates the mapping between continuous material properties and response functions by learning from their discretized vector representations. Each function is sampled over a fixed grid, enabling neural operator learning for metamaterial prediction.

# Tables

**Algorithm 1** MetaFO architecture

**Input:** Known material properties $\mathbf{x}_{\text{domain}} = \left\{ \mathbf{x}_{\text{domain}}^{(i)} \right\}_{i=1}^{k} = \left\{ \varepsilon_{m_i}, \sigma_{m_i}, \varepsilon_u \right\}_{i=1}^{k} \in \mathbb{R}^{T \times 3}$ solution features $\mathbf{x}_{\text{sol}} = \left\{ \mathbf{x}_{\text{sol}}^{(i)} \right\}_{i=1}^{k} = \left\{ \sigma_u^{m_i} \right\}_{i=1}^{k} \in \mathbb{R}^{T \times 1}$, $\mathbf{x}_{\text{domain}}^{\text{unknown}} = \{ \varepsilon_{m_{k+1}}, \sigma_{m_{k+1}}, \varepsilon_u \} \in \mathbb{R}^{T \times 3}$ prior noise $n_{\text{noise}}$, model parameters $\phi_{\text{domain}}, \phi_{\text{sol}}, \psi, \mathbf{E}_{\text{noise}}$, and constants $k, T, n, K, d_{\text{inp}}$.

**Output:** Predicted observable $Y = \sigma_u^{m_{k+1}} \in \mathbb{R}^{n \times T}$.

$$\mathbf{z}_{\text{domain}} = \phi_{\text{domain}}(\mathbf{x}_{\text{domain}}) \in \mathbb{R}^{k \times T \times d_{\text{inp}}},$$
$$\mathbf{z}_{\text{sol}} = \phi_{\text{sol}}(\mathbf{x}_{\text{sol}}) \in \mathbb{R}^{k \times T \times d_{\text{inp}}},$$
$$\mathbf{z} = \mathbf{z}_{\text{domain}} \oplus \mathbf{z}_{\text{sol}} \in \mathbb{R}^{k \times T \times 2d_{\text{inp}}},$$
$$\mathbf{E}_{\text{enc}} = \text{Flatten}(\mathbf{z}) \in \mathbb{R}^{(kT) \times 2d_{\text{inp}}},$$
$$\mathbf{H}_{\text{enc}} = \text{TransformerEncoder}(\mathbf{E}_{\text{enc}}) \in \mathbb{R}^{(nT) \times 2d_{\text{inp}}},$$
$$\mathbf{c}_{\text{global}} = \frac{1}{k} \sum_{i=1}^{k} \phi_{\text{sol}}\left(\mathbf{x}_{\text{sol}}^{(i)}\right) \in \mathbb{R}^{T \times d_{\text{inp}}},$$
$$\mathbf{w} = \psi\left(\text{Expand}(\mathbf{c}_{\text{global}})\right) \in \mathbb{R}^{(nT) \times K},$$
$$\mathbf{n}_{\text{dynamic}} = \mathbf{w} \cdot \mathbf{E}_{\text{noise}} \in \mathbb{R}^{(nT) \times d_{\text{inp}}},$$
$$\mathbf{E}_{\text{dec}} = \phi_{\text{domain}}\left(\mathbf{X}_{\text{domain}}^{\text{unknown}}\right) \oplus \mathbf{n}_{\text{dynamic}} \in \mathbb{R}^{(nT) \times 2d_{\text{inp}}},$$
$$\mathbf{H}_{\text{dec}} = \text{TransformerDecoder}(\mathbf{E}_{\text{dec}}, \mathbf{H}_{\text{enc}}) \in \mathbb{R}^{(nT) \times 2d_{\text{inp}}},$$
$$\mathbf{Y} = \text{MLP}(\mathbf{H}_{\text{dec}}) \in \mathbb{R}^{(nT)}.$$

# Supporting Information

## Toward a Robust and Generalizable Metamaterial Foundation Model

*Namjung Kim[*], Dongseok Lee, Jongbin Yu, Sung Woong Cho, Dosung Lee, Yesol Park, and Youngjoon Hong[*]*
* Corresponding authors

Jongbin Yu, Dosung Lee, Yesol Park, Prof. N. Kim
Department of Mechanical Engineering, Gachon University, Seongnam 13120, Republic of Korea

Dongseok Lee, Sung Woong Cho
Department of Mathematical Sciences, Korea Advanced Institute of Science and Technology, Daejeon, 34141, Republic of Korea

Prof. Y. Hong
Department of Mathematical Sciences, Seoul National University, Seoul, 08826, Republic of Korea
Interdisciplinary Program in Artificial Intelligence, Seoul National University, Seoul, 08826, Republic of Korea

*Corresponding author
Prof. Namjung Kim:          namjungk@gachon.ac.kr
Prof. Youngjoon Hong:       hongyj@kaist.ac.kr

## S1. Geometry of ten basis unitcells corresponding graph representations

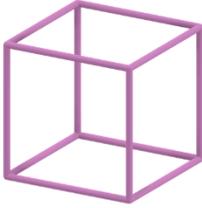

**Figure S1.** 10 basis unit cells were used to construct the initial database.

Graph-based representations were employed to design unitcells using a set of control nodes and their pairwise connectivity. This approach abstracts each unitcell into a network structure, where nodes represent spatial anchor points and edges define potential material connections. Figure S1 illustrates the geometry of ten distinct basis unitcells that form the foundation of the dataset. Each basis unitcell possesses uniqe, non-overlapping structural connectivity, allowing for linear combinations to create new configurations. Excluding the null case, this combinatorial strategy yiedls 1,023 unique unit cells derived from the ten basis structures. This graph-based design framework offers several key advantages for mechanical metamaterials: it enables efficient encoding of complex topologies, facilitates scalable exploration of desing spaces, and supports modular synthesis of new structures with tailored mechanical responses. Moreover, the ability to decouple and recombine basis structures provides a flexible platform for gnerating high-performance deisng sthat extend beyond convectional unitcell libraties.

## S2. Design space of base material properties and unitcell responses

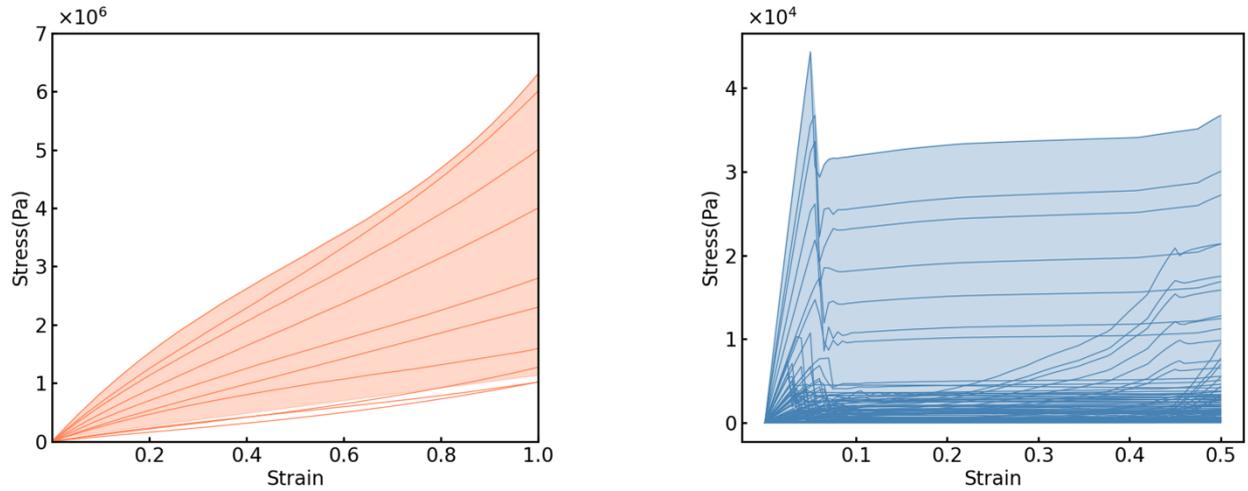

**Figure S2.** Design space of base materials and unitcell responses including experimental data

A hyperelastic model was employed to accurately represent various 3D-printable materials used in mechanical metamaterials. During numerical simulations involving large deformations, plasticity models often suffer from convergence issues due to irreversible strain accumulation after yielding. In contrast, hyperelastic models—formulated based on strain energy functions—provide superior numerical stability and robustness, making them particularly suitable for simulating the nonlinear mechanical responses of metamaterials. Additionally, hyperelastic models closely capture the behavior of elastomeric materials commonly used in additive manufacturing. To demonstrate manufacturability, three commercial elastomers from Formlabs—Elastic 50A, Flexible 80A, and Silicone 40A—were incorporated into the analysis. Seven virtual materials (denoted as Ex materials) were generated by interpolating the mechanical characteristics of these three base materials, thereby enabling a continuous and customizable range of material responses. The stress–strain curves, fitted using the Yeoh hyperelastic model, are presented in Figure S2. This modeling approach ensures both numerical reliability and practical relevance for soft material-based metamaterial design.

## S3. Numerical simulations for mechanical metamaterials under compression

The mechanical behavior of the designed unit cell structures under compressive loading was analyzed through finite element simulations conducted using a custom Python script interfaced with Abaqus/Explicit. Each unit cell was discretized using Timoshenko beam elements, making them suitable for modeling slender strut-like components commonly found in mechanical metamaterials. Each beam element consisted of two nodes to effectively capture the deformation mechanics of the microstructure.

To impose compression, rigid plates were introduced at the top and bottom of the unit cell. The bottom plate was constrained in all degrees of freedom (encastre boundary condition), while the top plate applied compressive displacement downward at a constant speed of 1 mm/min. This slow loading rate was chosen to suppress dynamic effects and approximate quasi-static conditions, a common practice in simulations where inertial contributions could otherwise distort the mechanical response.

A general contact algorithm was implemented to model interactions between different parts of the structure as they came into contact during deformation, which is critical in capturing mechanisms such as densification or strut contact. To further enhance contact fidelity, a kinematic contact formulation was used to limit artificial penetration of nodes into rigid bodies, which can occur in explicit solvers without proper constraint enforcement. The use of the explicit dynamic solver in Abaqus was motivated by its robustness in handling complex contact conditions and large deformations that arise in the nonlinear response of metamaterial structures. This setup allows for accurate prediction of the stress-strain response across various deformation regimes, including initial elastic behavior, buckling, post-yield plateau, and densification.

## S4. MetaFO's stress-strain curve predictions for various basis unitcell combinations

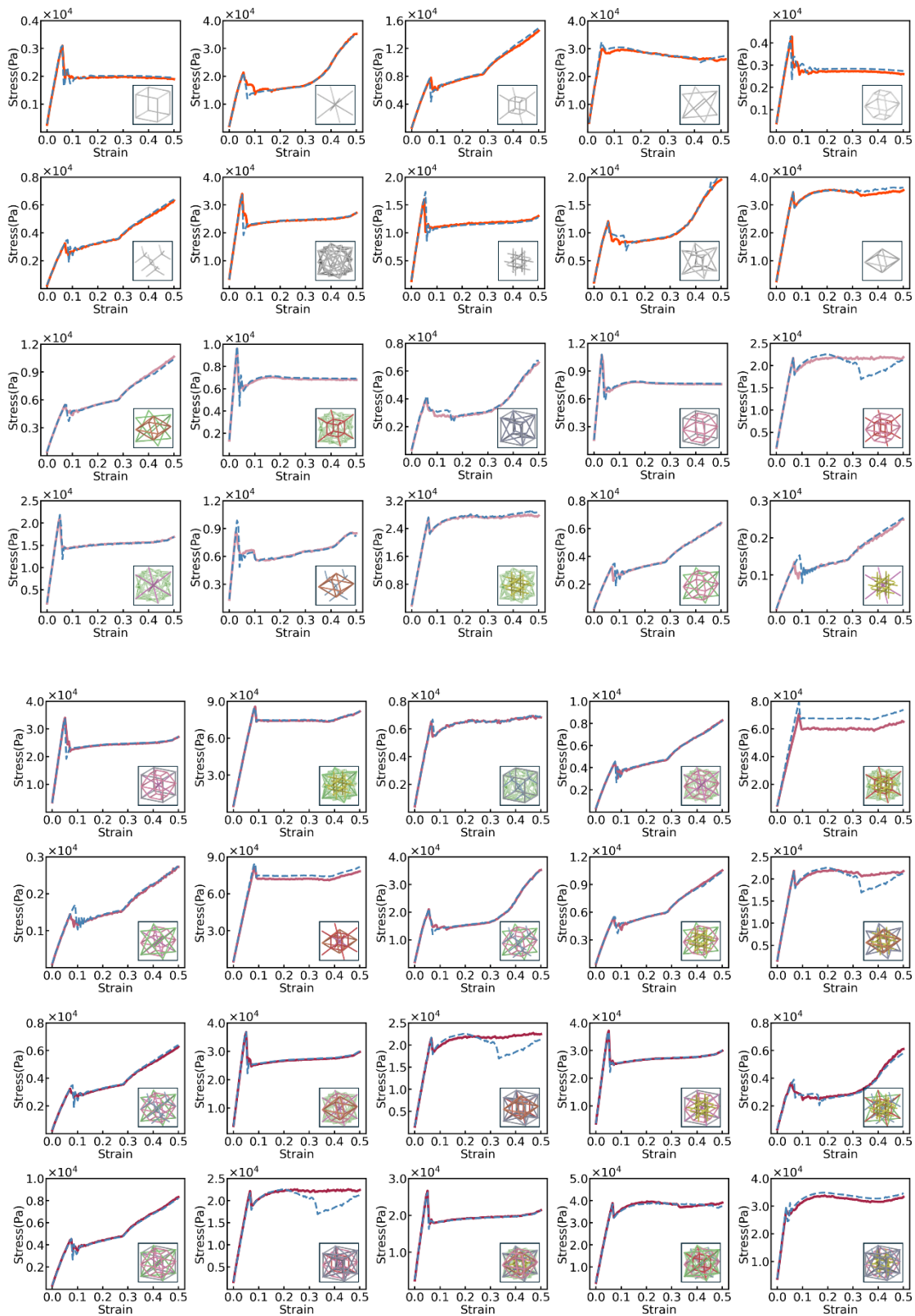

**Figure S3.** Compression test stress-strain curves for various unitcell combinations. All the combinations are four specimens are presented, highlighting the feasibility and reliability of the proposed modeling framework. A commercially available 3D printer, the Formlab 3, with UV-curable resin used to fabricate the samples. The fabrication parameters were rigorously defined to ensure the lattice structures were accurately manufactured. A 2x2x2 supercell was created to obtain the elastic modulus of the sample in the experimentally viable size. The compression tests were meticulously conducted using a universal testing machine (Technology & Dolf Ltd) equipped with a 10 kgf load cell, applying a small strain step to prevent dynamic effects on the specimens. The deformation patterns were captured and analyzed using the digital image correlation (DIC) method, providing insightful visualizations of the deformation responses.

## S5. Detailed architecutre for Transformer in MetaFO

MetaFO integrates dual encoders into a transformer-based encoder–decoder architecture to predict the metamaterial response function $f_u^m$ from the corresponding material properties $f_m$. This architecture is composed of three key components:

1. Dynamic noise context: A context-aware noise mechanism that adaptively generates noise embeddings conditioned on the given prompt. The noise weights are dynamically recomputed for every forward pass and for each stress evaluation, enabling fine-grained control based on input context.

2. Latent Representation Module: Neural modules that transform material and response function inputs into a shared high-dimensional latent space, allowing the model to jointly capture the structural relationship between material properties and their associated responses.

3. Cross-Attention Layer: A context-aware cross-attention module maps latent queries, derived from material property functions $f_m$, to their corresponding predicted responses. The Keys ($K$) and values ($V$) are obtained from known material–response pairs, which serve as contextual prompts.

We now provide a detailed mathematical formulation of these components. First, we introduce the dynamically generated noise context $\mathbf{n}_{\text{dynamic}} = \mathbf{w} \cdot \mathbf{E}_{\text{noise}} \in \mathbb{R}^{nT \times d_{\text{inp}}}$, computed as:

$$\mathbf{n}_{\text{dynamic},t} = \mathbf{w}_t \cdot \mathbf{E}_{\text{noise}} \in \mathbb{R}^{n \times d_{\text{inp}}}, \qquad t = 1, \dots, T,$$

where the coefficient $w_t \in \mathbb{R}^{n \times K}$ are given by:

$$w_t = \psi\left(\frac{1}{k}\sum_{i=1}^{k} \phi_{\text{sol}}\left(\mathbf{x}_{\text{sol},t}^{(i)}\right)\right) \in \mathbb{R}^{n \times K}.$$

Here, the parameters of $\psi$ and $\phi_{\text{sol}}$ are fixed after training, while the $K$ coefficients $w_t$ vary with each new prompt $\mathbf{x}_{\text{sol},t}^{(i)}$ during inference. Consequently, the dynamically generated noise $\mathbf{n}_{\text{dynamic}}$ adapts to each prompt. Moreover, if different types of noise are imposed on $\mathbf{x}_{\text{sol},t}^{(i)}$, the corresponding $\mathbf{n}_{\text{dynamic}}$ changes accordingly, reflecting the variations induced by these distinct noise conditions.

Second, we detail the transformer encoder step. Consider the encoder as a stack of $L$ layers with identical hidden units. Starting with $X^{(0)} = \mathbf{E}_{\text{enc}}$, we iterate the following transformations for each layer $l = 1, \dots, L$.

$$Z^{(l)} = \text{LN}\left(X^{(l-1)} + \text{MHA}^{(l)}\left(X^{(l-1)}\right)\right),$$

$$X^{(l)} = \text{LN}\left(Z^{(l)} + \text{FFN}^{(l)}\left(Z^{(l)}\right)\right).$$

Here, $\text{MHA}^{(l)}$ represents the multi head attention mechanism, defined as:

$$\text{MHA}^{(l)}(X) = \left[\text{head}_1^{(l)} \| \dots \| \text{head}_H^{(l)}\right] W_O^{(l)} \in \mathbb{R}^{kT \times 2d_{\text{inp}}},$$

where each head is computed as:

$$\text{head}_h^{(l)} = \text{softmax}\left(\frac{Q_h K_h^\top}{\sqrt{d_k}}\right) V_h, Q_h = X W_{Q,h}^{(l)}, K_h = X W_{K,h}^{(l)}, V_h = X W_{V,h}^{(l)}.$$

Layer Normalization (LN) is constructed as follows with the trainable parameter $\gamma, \beta \in \mathbb{R}^{2d_{\text{inp}}}$.

$$LN(x) = \gamma \odot (x - \mu(x))/\sqrt{\sigma^2(x) + 10^{-5}} + \beta.$$

Note that the position-wise feed-forward block is given by:

$$\text{FFN}^{(l)}(Z) = \sigma(ZW_1^{(l)} + b_1^{(l)})W_2^{(l)} + b_2^{(l)},$$

with GELU as the activation function $\sigma$. The final encoder memory is thus:

$$\mathbf{H}_{\text{enc}} = X^{(L)} \in \mathbb{R}^{kT \times 2d_{\text{inp}}}$$

Finally, we describe the transformer decoder block, which incorporates a cross-attention mechanism and leverages encoder-generated memory. Initially, the decoder state is set as $U^{(0)} = \mathbf{E}_{\text{dec}}$. Each subsequent layer $l$ recursively computes:

$$\widetilde{U}^{(l)} = \text{LN}\left(U^{(l-1)} + \text{MHA}_{\text{self}}^{(l)}(U^{(l-1)})\right),$$

$$\widehat{U}^{(l)} = \text{LN}\left(\widetilde{U}^{(l)} + \text{MHA}_{\text{cross}}^{(l)}(\widetilde{U}^{(l)}, \mathbf{H}_{\text{enc}})\right),$$

$$U^{(l)} = \text{LN}(\widehat{U}^{(l)} + \text{FFN}^{(L+l)}(\widehat{U}^{(l)})).$$

The cross-attention mechanism utilizes queries derived from the decoder and keys and values from the encoder memory $H_{enc}$, computed as follows:

$$\text{MHA}_{\text{cross}}^{(l)}(Q; K, V) = [\text{head}_1^{(l)} \| \ldots \| \text{head}_H^{(l)}]W_O^{(L+l)}$$

$$\text{head}_h^{(l)} = \text{softmax}\left(\frac{Q_h K_h^\top}{\sqrt{d_k}}\right)V_h, \quad Q_h = \widetilde{U}^{(l)}W_{Q,h}^{(d,l)}, K_h = H_{\text{enc}} W_{K,h}^{(e,l)}, V_h = H_{\text{enc}} W_{V,h}^{(e,l)}$$

Finally, we can get the following output.

$$\mathbf{H}_{\text{dec}} = U^{(L)} \in \mathbb{R}^{nT \times 2d_{\text{inp}}}$$

Throughout this paper for all tasks, we set the number of attention heads $H = 4$, number of layers $L = 3$, input embedding dimension $d_{\text{inp}}=64$, and uniformly define all hidden layer dimensions across the network as 128.

By integrating these components, MetaFO effectively captures the complicated operator mappings between material functions and their corresponding metamaterial responses. The model design manages uncertainties such as noise and incomplete data through context-conditioned noise injection. Furthermore, MetaFO leverages prompt-based, multi-head cross-attention mechanism, enabling flexible and context-aware predictions. This ability supports both forward and inverse design tasks in metamaterial engineering with high precision.

### S6. Detailed training procedure for MetaFO

To train the MetaFO model, we first prepare a comprehensive dataset. Specifically, we consider multiple unit cell geometries $\{u_j\}_{j=1}^{U}$, material types $\{m_i\}_{i=1}^{M}$, and discretized strain value $\{\varepsilon_{u_j,t}\}_{t=1}^{T}$ for each unit cell geometry $u_j$. Through simulations, we generate discretized material property values $\{f_{m_i}(\varepsilon_{m_i,t}) = \sigma_{m_i,t}\}_{t=1}^{T}$ for each material $m_i$, and corresponding metamaterial response functions $\{f_{u_j}^{m_i}(\varepsilon_{u_j,t}) = \sigma_{u_j,t}^{m_i}\}_{i=1,\ldots,M, j=1,\ldots,U, t=1,\ldots,T}$. For training, the model utilizes a prompt consisting of $k$ known pairs $\left(f_{m_i}(\varepsilon_{m_i,t}), f_{u_j}^{m_i}(\varepsilon_{u_j,t})\right)$ drawn from selected indices $i$. Given an input of material property functions $f_{m_*}(\varepsilon_{m_*,t})$ for a new material $m_*$, the model predicts corresponding metamaterial response functions $f_{u_j}^{m_*}(\varepsilon_{u_j,t})$. Finally, the dataset indices $[U] \coloneqq \{1, 2, \ldots, U\}$ and $[M] \coloneqq \{1, 2, \ldots, M\}$ are partitioned into training and test subsets $U_{train}, U_{test}$ and $M_{train}, M_{test}$ with an 80:20 ratio, respectively. MetaFO is trained using only data indexed by $j \in U_{train}$ and $i \in M_{train}$, while evaluation is performed on unseen materials and unit cell geometries from the test sets.

MetaFO is trained by minimizing the discrepancy between predicted and true response functions across diverse material-property distributions. We used empirical risk minimization with the mean squared error (MSE) loss function:

$$\mathcal{L}(\boldsymbol{\theta}) = \frac{1}{UMT} \sum_{j=1}^{U} \sum_{i=1}^{M} \sum_{t=1}^{T} \left\| f_{u_j}^{m_i}\left(\varepsilon_{u_j,t}\right) - \hat{f}_{u_j}^{m_i}\left(\varepsilon_{u_j,t}; \boldsymbol{\theta}\right) \right\|^2,$$

where $f_{u_j}^{m_i}\left(\varepsilon_{u_j,t}\right)$ is the true response function and $\hat{f}_{u_j}^{m_i}\left(\varepsilon_{u_j,t}; \boldsymbol{\theta}\right)$ is the predicted response function. To minimize $\mathcal{L}(\boldsymbol{\theta})$, we iteratively update parameters $\boldsymbol{\theta}$. The overall training procedure is summarized as follows:

---

**Algorithm 1** MetaFO Training

---

**Input:** Dataset $D = \left\{ f_{m_i}(\varepsilon_{m_i,t}), f_{u_j}^{m_i}\left(\varepsilon_{u_j,t}\right) \right\}_{i \in M_{train}, j \in U_{train}, t \in \{1,\dots,T\}}$, Initial parameters $\theta$, Batch size $B$, Number of iterations $T$

For epoch=1 to $N$ do:
    Sample a batch $B \subseteq D$ of $B$ function pairs.
    Construct a prompt $\mathcal{P}$ from $B$ (select $k$ known pairs with shared $u_i$).
    Select a query material property $f_{m_i}$ from $B$.
    Compute the predicted response function $\hat{f}_{u_j}^{m_i}\left(\varepsilon_{u_j,t}; \boldsymbol{\theta}\right)$
    Calculate the loss:
$$\mathcal{L}(\boldsymbol{\theta}) = \frac{1}{UMT} \sum_{j=1}^{U} \sum_{i=1}^{M} \sum_{t=1}^{T} \left\| f_{u_j}^{m_i}\left(\varepsilon_{u_j,t}\right) - \hat{f}_{u_j}^{m_i}\left(\varepsilon_{u_j,t}; \boldsymbol{\theta}\right) \right\|^2,$$
    Update parameters $\boldsymbol{\theta}$ using gradient descent:
        $\boldsymbol{\theta} \leftarrow \boldsymbol{\theta} - \eta \nabla_{\boldsymbol{\theta}} \mathcal{L}(\boldsymbol{\theta})$, where $\eta$ is the learning rate.

**Output:** Trained model parameters $\boldsymbol{\theta}$.

---

The training procedure also incorporates structured noise injection, where response functions are perturbed by adding noise sampled from a learnable noise bank. This noise is conditioned on the global context to reflect realistic variations in material-response behavior. This process prevents overfitting to specific training examples and encourages the model to capture generalizable patterns, allowing it to recover physically consistent predictions even in the presence of uncertainties or missing data.

## S7. Inverse design capability of MetaFO

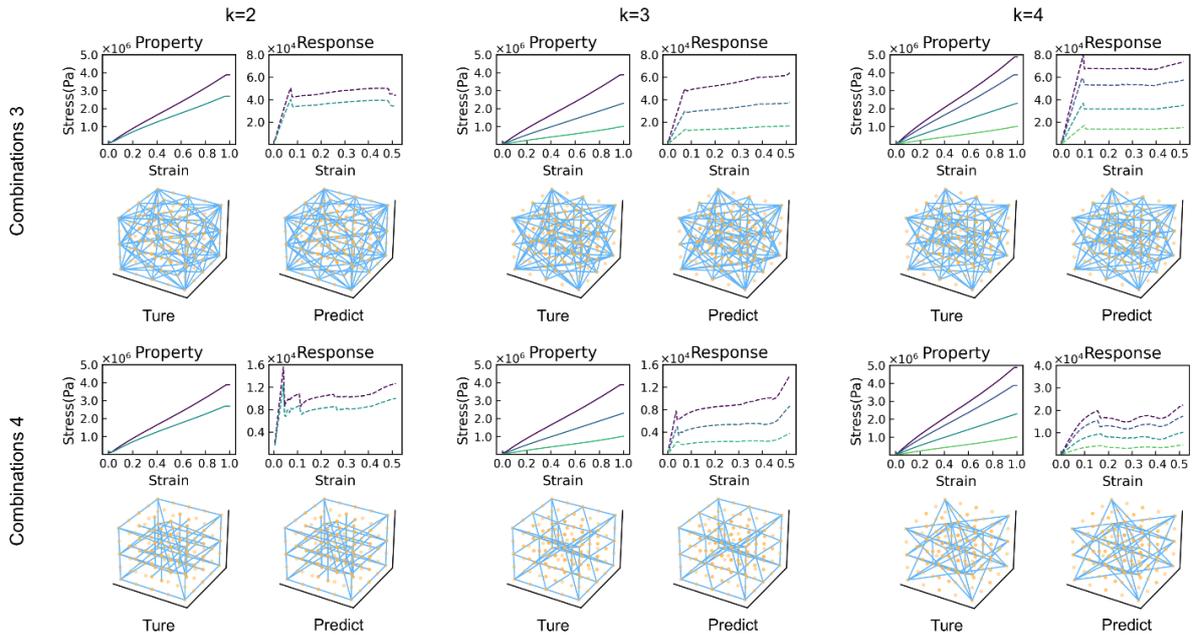